\documentclass{article}
\usepackage{ijcai23}

\usepackage{times}
\usepackage{soul}
\usepackage{url}
\usepackage[hidelinks]{hyperref}
\usepackage[utf8]{inputenc}
\usepackage[small]{caption}
\usepackage{graphicx}
\usepackage{amsmath}
\usepackage{amsthm}
\usepackage{booktabs}
\usepackage{algorithm}
\usepackage{algorithmic}
\usepackage[switch]{lineno}
\usepackage{amsfonts}
\usepackage{bm} 
\usepackage{subfigure}
\usepackage{enumitem}
\usepackage[dvipsnames]{xcolor}

\title{Full Scaling Automation for Sustainable Development of Green Data Centers}

\author{
    Shiyu Wang, Yinbo Sun, Xiaoming Shi, Zhu Shiyi, Lin-Tao Ma, James Zhang, 
     \vspace{1mm} YangFei Zheng \And  Liu Jian
    \affiliations
    Ant Group
    \emails
    \{weiming.wsy, yinbo.syb, peter.sxm, zhushiyi.zsy, lintao.mlt, james.z, yangfei.zyf, basti.lj\}@antgroup.com
}

\begin{document}

\maketitle

\begin{abstract}
The rapid rise in cloud computing has resulted in an alarming increase in data centers' carbon emissions, which now account for $>$3\% of global greenhouse gas emissions, necessitating immediate steps to combat their mounting strain on the global climate.
An important focus of this effort is to improve resource utilization in order to save electricity usage.  
Our proposed Full Scaling Automation (FSA) mechanism is an effective method of dynamically adapting resources to accommodate changing workloads in large-scale cloud computing clusters, enabling the clusters in data centers to maintain their desired CPU utilization target and thus improve energy efficiency.
 FSA harnesses the power of deep representation learning to accurately predict the future workload of each service and automatically stabilize the corresponding target CPU usage level, unlike the previous autoscaling methods, such as Autopilot or FIRM, that need to adjust computing resources with statistical models and expert knowledge.  Our approach achieves significant performance improvement compared to the existing work in real-world datasets.  We also deployed FSA on large-scale cloud computing clusters in industrial data centers, and according to the certification of the China Environmental United Certification Center (CEC), a reduction of 947 tons of carbon dioxide, equivalent to a saving of 1538,000 kWh of electricity, was achieved during the Double 11 shopping festival of 2022, marking a critical step for our company’s strategic goal towards carbon neutrality by 2030. 


\end{abstract}

\section{Introduction}
As the demand for data and cloud computing services continues to soar, the data center industry is expanding at a staggering rate, serving as the backbone of the economic, commercial, and social lives of the modern world.  The data centers are, however, some of the world’s biggest consumers of electrical energy, and their ever-increasing carbon emissions only exacerbate global warming. 
Currently, the emissions from data centers compose 3.7\% of all global greenhouse gas emissions, exceeding those from commercial flights and other existential activities that fuel our economy.  This attracts the concern of many global organizations, including the United Nations' climate conferences focusing on sustainability in data centers. 
 As part of the worldwide effort towards \textit{carbon neutrality} (or \textit{net zero carbon}), making more efficient use of data centers' computing resources becomes an increasingly popular research area to minimize the energy consumption. 

The carbon emissions of large-scale cloud computing clusters in data centers mainly come from the power consumption of equipment loads. Several studies have shown that servers running with chronically low CPU utilization are one major source of energy wasting, which can be alleviated by reducing the amount of resources allocated for low-load services, and reallocating those saved servers to high-load services\cite{krieger2017building,abdullah2020burst}. 
 Efficient resource allocation mechanism thus serves an important role in this effort.  Moreover, some systems allow flexibly shutting down unneeded servers to further save the power consumption with proper scheduling and engineering mechanism.
To this extent, most cloud computing providers try to take advantage of automatic scaling systems (autoscaling) to stabilize the CPU utilization of the systems being provisioned to the desired target levels, not only to ensure their services meet the stringent Service Level Objectives (SLOs), but also to prevent CPUs from running at low utilization\cite{qiu2020firm}.

Specifically, autoscaling elastically scales the resources horizontally (i.e., changing the number of virtual Containers (Pods) assigned) or vertically (i.e., adjusting the CPU and memory reservations) to match the varying workload according to certain performance measure, such as resource utility.  Prior work on cloud autoscaling can be categorized as rule-based or learning-based schemes. 
Rule-based mechanism sets thresholds of metrics (such as CPU utilization and workload ) based on expert experience and monitors indicator changes to facilitate resource scheduling. Learning-based autoscaling mechanism employs statistical or machine learning to model historical patterns to generate scaling strategies. However, the existing works face the following challenges: 
\begin{itemize}[leftmargin=*]
\item \textbf{Challenge 1}: \textbf{Forecasting \textit{time series} (TS) of workloads with complex periodic and long temporal dependencies.} Most autoscaling studies focus on decision-making processes of scaling rather than server workload forecasting, eluding the critical task of accurately predicting future workloads (normally non-stationary TS of high-frequency with various periods and complex long temporal dependencies), which are not well-handled by existing work employing classic statistical methods (such as ARIMA) or simple neural networks (such as RNN). 

\item \textbf{Challenge 2}: \textbf{Maintaining stable
CPU utilization and characterizing uncertainty in the decision-making process.} Most methods aim at adequately utilizing resources for cost-savings while ignoring the stability measures of the service, such as the robustness and SLOs assurances, where stable CPU utilization is of great significance.  Moreover, affected by various aspects of the servers (such as temperature), the uncertainty of the correlation between CPU utilization and workload needs to be modeled in the decision-making process to balance between benefits and risks.

\item \textbf{Challenge 3}: \textbf{Accounting for sustainable development of green data centers.} Existing works only optimize resources and costs, instead of carbon emissions, which is one essential task for the sustainable green data centers.
\end{itemize}

\noindent To this extent, we propose in this paper a novel predictive horizontal autoscaling named Full Scaling Automation (FSA) to tackle the above difficulties.
Specifically, \textbf{first}, we develop an original workload forecasting framework based on representation learning, i.e., we learn multi-scale representations from historical workload TS as external memory, characterizing long temporal dependencies and complex various periodic (such as days, weeks and months).  We then build a representation-enhanced deep autoregressive model, which integrates multi-scale representations and near-observation TS via multi-head self-attention. 
\textbf{Second}, we propose a task-conditioned Bayesian neural model to learn the relationship between CPU utilization and workload, enabling the characterization of the uncertainty between them, while providing the upper and lower bounds of benefits and risks. Moreover, with the help of the task-conditioned hypernetwork \cite{mahabadi2021parameter,he2022hyperprompt}, our approach can be easily adapted to various online scenarios. \textbf{Third}, based on the workload forecasting and CPU utilization estimation above, we can properly pre-allocate resources and distribute the future workload to each machine, targeting certain CPU utilization level, maximizing energy efficiency.

We have deployed our FSA mechanism on large-scale cloud computing clusters, and with the help of China Environmental United Certification Center (CEC), one can convert between energy saving and carbon emission reduction, aiming at sustainable development in data centers. To the best of our knowledge, we are the first in the industry to employ AI-based resource scaling for this goal.

\textbf{Our Contributions} can be summarized as follows:
\begin{itemize}[leftmargin=*]
 \item A novel workload time series forecast framework with the representation-enhanced deep autoregressive model, integrating multi-scale TS representation.
\item A task-conditioned Bayesian neural model estimating the uncertainty between CPU utilization and workload.
\item An efficient predictive horizontal autoscaling mechanism to stabilize CPU utilization, saving costs and energy, facilitating sustainable data centers.
\end{itemize}

\begin{figure*}[h]
  \centering
  \includegraphics[width=\linewidth, height=0.5\linewidth]{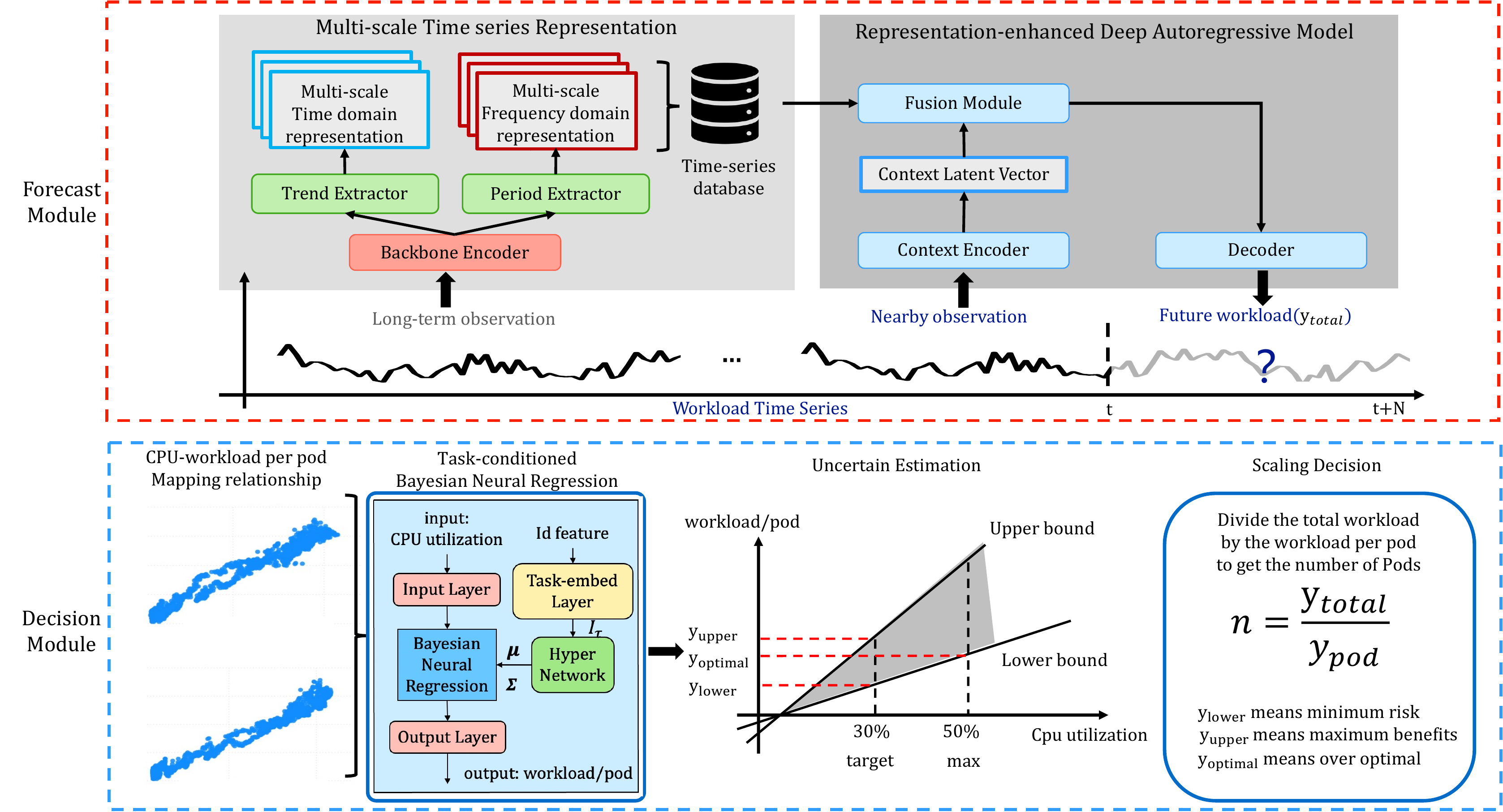}
  \caption{\textbf{Model Architecture.} Red dashed lines represent the workload forecast module, including Multi-scale Time Series Representation and Representation-enhanced Deep Autoregressive Model. Blue dashed lines highlight the scaling decision module via task-conditioned Bayesian neural regression. According to service tolerance for \textit{response time(rt)}, the optimal value between upper and lower bounds is obtained.}
  \label{fig:fig1}
\end{figure*}

\section{Background And Related Work}
\subsection{Horizontal Autoscaling}
Horizontal autoscaling is a practical resource management method that elastically scales the resources horizontally (i.e., changing the number of virtual Containers (Pods) assigned) \cite{nguyen2020horizontal}.  Most existing studies focus on avoiding service anomalies, e.g., FIRM \cite{qiu2020firm} uses machine learning to detect service performance anomalies (e.g., the RT of a microservice is abnormally long) and when such anomalies occur, more Pods are allocated;
Autopilot \cite{rzadca2020autopilot} takes TS of the CPU utilization of a microservice as input, and employs a simple heuristic mechanism to obtain the target CPU utilization, which is then used to calculate the number of Pods required as a linear function. Pods are increased or reduced to minimize the risk that a microservice suffers from anomalies. These methods have the following limitations: (1) Most methods are reactive scaling, i.e., autoscaling occurs only after performance anomalies occur. (2) Due to the lack of workload forecasting mechanism, making precise decision for future resource scaling is difficult. (3) The uncertainty of CPU utilization is not considered in decision making, leading to unknown risks in resource scaling.

\subsection{Workload Forecasting}
One key constituent of our method is forecasting the workload, which is, however, normally of high-frequency and non-stationary with various periods and complex long temporal dependencies, defying most classic TS forecasting models.  We, therefore, propose to use TS representations to memorize long-term temporal dependencies to battle this complexity.

\paragraph{Time Series (TS) Forecasting.}
Deep forecasting methods based on RNN or Transformer, including N-beats, Informer, and DeepAR, etc., have been widely applied to TS forecasting, outperforming classical models such as ARIMA and VAR.  
N-BEATS is an interpretable TS forecasting model, with deep neural architecture based on backward and forward residual links with a very deep stack of fully-connected layers\cite{oreshkin2019n}.
Informer is a prob-sparse self-attention mechanism-based model to enhance the prediction capacity in the long-sequence TS forecasting problems, which validates the Transformer-like model’s potential value to capture individual long-range dependency between long sequence TS outputs and inputs \cite{zhou2021informer}.
DeepAR is a probabilistic autoregressive forecasting model based on RNN, widely applied in real-world industrial applications \cite{salinas2020deepar}.

\paragraph{Time Series (TS) Representation.}
Representation learning has recently achieved great success in advancing TS research by characterizing the long temporal dependencies and complex periodicity based on the contrastive method. 
TS-TCC creates two views for each sample by applying strong and weak augmentations and then using the temporal contrasting module to learn robust temporal features by applying a cross-view prediction task \cite{eldele2021timetstcc}.
TS2Vec was recently proposed as a universal framework for learning TS representations by performing contrastive learning in a hierarchical loss over augmented context views \cite{yue2022ts2vec}. 
A new TS representation learning framework was proposed in CoST for long-sequence TS forecasting, which applies contrastive learning methods to learn disentangled seasonal-trend representations \cite{woo2022cost}.

\subsection{Bayesian Method}
Bayesian methods are frequently used to model uncertainties of the variables, and are introduced to deep learning models to improve robustness of the model \cite{bishop2006pattern}.  In this work, we introduce task-conditioned hypernetwork \cite{mahabadi2021parameter} into Bayesian regression to estimate the correlation between CPU utilization and workload.

\section{Proposed Methodology}
In this section, we detail our framework, including the forecast module and decision module, as shown in Figure 1.

\subsection{Problem Definition}
Let us denote the value of TS of workload by $y_{t} \in \mathbb{R}^{F}$, where $t$ indexes time on a time horizon $t\in \{1,2,..., T\}$, and $F$ is the dimension of features, such as time feature and id feature.

Given the historical workload TS of an application $y_{t-L}$ at time $t$, we aim to find the optimal Pods allocated over the future time horizon $N$ to make the application run stably around a target CPU utilization. We assume the total workload of an application is evenly allocated across all Pods via load balancing and we can represent their relationship as
\begin{equation}
\label{eq:eq1}
y^{pod}_{t}={y^{total}_{t}}/{n_{t}} 
\quad \Leftrightarrow \quad 
n_{t}={y^{total}_{t}}/{y^{pod}_{t}}
\mbox{ ,}
\end{equation}
where $y^{total}_{t}$ is the total workload, $y^{pod}_{t}$ is the workload per Pod, and $n_{t}$ is the number of Pods at time $t$.  Please note that this assumption is ensured by our software and hardware engineering platform of the data centers.

\subsection{Workload Forecasting}
Considering that the workload TS are non-stationary and of high-frequency (one minute scale) with various periods and complex long temporal dependencies (such as daily, weekly, monthly, and even quarterly periodicities) and existing work requires backcasts on long historical time windows as the input of the context.  The formidable amount of data makes memorizing the historical data and learning these temporal dependencies difficult.  Therefore, we propose a TS time series representation method, which characterizes complex long-term historical TS with various periodicity as compressed representation vectors and stores them in the TS database.  We then design a deep autoregressive fusing model, which integrates long-term historical TS representations and short-term observations (nearby window) to achieve accurate predictions.

\subsubsection{Multi-scale Time Series (TS) Representation}
Given TS $y \in \mathbb{R}^{T \times F}$ with backcast window $h$, our goal is to learn a nonlinear embedding function $f_{\theta
}$ that maps $\{y_{t-h}...y_{t}\}$ to its representation $r_{t}=[r^{T}_{t},r^{F}_{t}]$, where $r_{t} \in \mathbb{R}^{K}$ is for each time stamp $t$, $r^{T}_{t} \in \mathbb{R}^{K_{T}}$ is the time domain representation, $r^{F}_{t} \in \mathbb{R}^{K_{F}}$ denotes that of frequency domain and $K=K_{T}+K_{F}$ is the dimension of representation vectors.  In the encoding representation stage, by using backcast windows of various lengths, we can obtain a representation of different scales.

 We first randomly sample two overlapping subseries from an input TS and then perform data augmentation separately.  Next, using Multilayer Perceptron (MLP) as the input projection layer, the original input ${y_{t}}$ is mapped into a high-dimensional latent vector ${z_{t}}$.  We use timestamp masking to mask latent vectors at randomly selected timestamps to generate an augmented context view.  We then use CovnTrans as the backbone encoder to extract the contextual embedding at each timestamp. Subsequently, we extract the trends in the time domain and the period in the frequency domain using CausalConv and Fast Fourier transform (FFT) from the contextual embedding, respectively. Finally, we perform contrastive learning in the time and frequency domains.  
 In the following sections, we describe these components in detail.

\textbf{Random Cropping} is a data augmentation method commonly used in contrastive learning to generate new context views. We can randomly sample two overlapping time segments $[a_1, a_2]$ and $[b_1, b_2]$ from TS $y \in \mathbb{R}^{T \times F}$ that satisfy $0<a_1<b_1<a_2<b_2$ $\le$ $T$. Note that contextual representations on the overlapped segment $[b1, a2]$ ensure consistency for two context views.

\textbf{Timestamp Masking} aims to produce an augmented context view by randomly masking the timestamps of a TS. We can mask off the latent vector $z=\{z_{t}\}$  after the Input Projection Layer along the time axis with a binary mask $m \in \{0, 1\}^T$, the elements of which are independently sampled from a Bernoulli distribution with $p = 0.5$. 

\textbf{Backbone Encoder} is used to extract the contextual representation at each timestamp. We use 1-layer causal convolution Transformer (ConvTrans) as our backbone encoder, which is enabled by its convolutional multi-head self-attention to capture both long- and short-term dependencies.

Specifically, given TS $y \in \mathbb{R}^{T \times F}$, ConvTrans transforms $y$ (as input) into dimension $l$ via dilated causal convolution layer as follows:
\begin{equation*}
\begin{aligned}
Q&=Dilated Conv(y) \\
K&=Dilated Conv(y) \\
V&=Dilated Conv(y) 
\end{aligned}\mbox{ ,}
\end{equation*}
where $\bm{Q} \in \mathbb{R}^{dl \times dh}$, $\bm{K} \in \mathbb{R}^{dl \times dh}$, and $\bm{V} \in \mathbb{R}^{dl \times dh}$ (we denote the length of time steps as $dh$). 
After these transformations, the scaled dot-product attention computes the sequence of vector outputs via:
\begin{equation*}
    {S}=\text{Attention}({Q},\ {K},\ {V})=\operatorname{softmax}\left({{Q} {K}^{T}}/{\sqrt{d_{K}}} \cdot {M}\right ) {V}
 \mbox{ ,}
 \end{equation*}
 where the mask matrix $M$ can be applied to filter out right-ward attention (or future information leakage) by setting its upper-triangular elements to $-\infty$ and normalization factor $d_{K}$ is the dimension of $W_{h}^{K}$ matrix.  Finally, all outputs $S$ are concatenated and linearly projected again into the next layer. After the above series of operations, we use this backbone $f_{\theta}$ to extract the contextual embedding (intermediate representations ) at each timestamp as $\tilde{r}=f_{\theta}(y)$.

\paragraph{Time Domain Contrastive Learning.} To extract the underlying trend of TS, one straightforward method is to use a set of 1d causal convolution layers (CasualConv) with different kernel sizes and an average-pooling operation to extract the representations as
\begin{equation*}
\tilde{r}^{(T,i)}=CausalConv(\tilde{r},2^i) 
\end{equation*}
\begin{equation*}
r^{T}=AvgPool(\tilde{r}^{(T,1)},\tilde{r}^{(T,2)}...,\tilde{r}^{(T,L)}) \mbox{ ,}
\end{equation*}
where $L$ is a hyper-parameter denoting the number of CasualConv, $2^i$ ($i=0,..., L$) is the  kernel size of each  CasualConv, $\tilde{r}$ is above intermediate representations from the backbone encoder, followed by average-pool over the $L$ representations to obtain time domain representation $\tilde{r}^{(T)}$. To learn discriminative representations over time, we use the time domain contrastive loss, which takes the representations at the same timestamp from two views of the input TS as positive samples ($r^{T}_{i,t},\hat{r}^{T}_{i,t}$), while those at different timestamps from the same time series as negative samples, formulated as
\begin{small}
\begin{equation*}
\mathcal{L}_{time}=-log\frac{exp(r^{T}_{i,t}\cdot\hat{r}^{T}_{i,t})}
{\sum_{t^{\prime} \in \mathcal{T}}(exp(r^{T}_{i,t}\cdot\hat{r}^{T}_{i,t^{\prime}})+\mathbb{I}(t \ne t^{\prime})exp(r^{T}_{i,t}\cdot r^{T}_{i,t^{\prime}})}
\end{equation*}
\end{small}
, where $\mathcal{T}$ is the set of timestamps within the overlap of the two subseries, subscript $i$ is the index of the input TS sample, and $t$ is the timestamp.

\paragraph{Frequency Domain Contrastive Learning.} Considering that spectral analysis in the frequency domain has been widely used in period detection, we use Fast Fourier Transforms (FFT) to map the above intermediate representations to the frequency domain to capture different periodic patterns. We can thus build a period extractor, including FFT and MLP, which extracts the frequency spectrum from the above contextual embedding and maps it to the freq-based representation $r^{F}_{t}$.
To learn representations that are able to discriminate between different periodic patterns, we adopt the frequency domain contrastive loss indexed with $(i,t)$  along the instances of the batch, formulated as
\begin{small}
\begin{equation*}
\mathcal{L}_{Freq}=-log
\frac{exp(r^{F}_{i,t}\cdot\hat{r}^{F}_{i,t})}
{\sum_{j \in \mathcal{D}}(exp(r^{F}_{i,t}\cdot\hat{r}^{F}_{i,t^{\prime}})+\mathbb{I}(i \ne j)exp(r^{F}_{i,t}\cdot r^{F}_{i,t^{\prime}})}
\end{equation*}
\end{small}
, where $\mathcal{D}$ is defined as a batch of TS. We use freq-based representations of other TS at timestamp $t$ in the same batch as negative samples.

The contrastive loss is composed of two losses that are complementary to each other and is defined as
\begin{equation}
\begin{aligned}
      \mathcal{L}=
      \frac{1}{|\mathcal D|T}(\mathcal{L}_{time}+\mathcal{L}_{Freq})
\end{aligned}
 \mbox{ ,}\end{equation}
where $\mathcal D$ denotes a batch of TS.
As mentioned before, we pre-train our TS representation model, and in the encoding representation stage, by using backcast windows of various lengths, we can obtain a representation of different scales. In this paper, we encode the origin high-frequency TS(one-minute scale) to generate daily, weekly, monthly, and quarterly representations of long-term historical TS.

\subsubsection{Representation-enhanced Deep Autoregressive Model}
We now introduce the Representation-enhanced Deep Autoregressive Model in detail. Let us denote the values of TS by $y_{t}\in \mathbb{R}$, where $t$  indexes time on a time horizon $t\in \{1,2,...,T\}$.
Note that we define $x_{t}$ as future knowable covariates, such as time feature and id feature, at time step $t$.
Given the last $L$ observations $y_{t-L},...,y_{t}$, the TS forecasting task aims to predict the future N observations $y_{t+1},...,y_{t+N}$. 
We use $r_t$, the representation of the long-term historical TS, and $h_{t}$ as the context latent vector from short-term observations (nearby window), to predict future observations. 

Specifically, we first load the Multi-scale TS representation $r_{t}$ (including daily, weekly, monthly, and even quarterly) from the TS database, which characterizes various periods and complex long temporal dependencies. Secondly, we use the Recurrent Neural Network (RNN) to encode the short-term observations (nearby window) as a context latent vector $h_{t}$, which captures the intraday changing pattern of workload TS. 
We then use MLP to integrate them into the same dimension. Thirdly, inspired by the transformer architecture, we naturally use the multi-head self-attention as the fusion module that integrates long-term historical TS representations and short-term observations (nearby window) to achieve better predictions. This architecture enables us to capture not only the long-term dependency and periods of TS, but also the short-term changes within the day. Finally, with the help of RNN as the decoder, we perform autoregressive decoding to achieve multi-step prediction.

In the training, given $\mathcal D$, defined as a batch of TS $Y:=\{y_{1},y_{2},...,y_{T}\}$, the representation of the long-term historical TS as $r_{t}$, 
and the associated covariates $X:=\{x_{1},x_{2},...,x_{T}\}$, we can derive the likelihood as:
\begin{equation}
\begin{aligned}
      \mathcal{L}=
      \frac{1}{|\mathcal D|T}\prod_{x1:T,y1:T \in \mathcal D}\prod_{t=1}^{T} p(y_{t}|y_{1:t-1};x_{1:t}, r_{t},\Phi) 
\end{aligned}
 \mbox{ ,}\end{equation}
 where $\Phi$ is the learnable parameter of the model.

\subsection{Scaling Decision}
As mentioned before, we aim to find the optimal Pods allocated over the future time horizon $N$ to make the application run stably around a target CPU utilization, with the assumption that the total workload of an application is evenly allocated on all Pods by load balancing, i.e., the total future workload $y_{t}^{total}$ is evenly distributed over all Pods. Therefore, in order to obtain the number of scaling Pods, according to equation~\ref{eq:eq1} above, our current key is to learn the relationship between CPU utilization and workload, for which, we propose Task-conditioned Bayesian Neural Regression.

\paragraph{Task-conditioned Bayesian Neural Regression.}
 We propose a novel method that incorporates the task-conditioned hypernetwork and the Bayesian method to create a unified model that can adapt to varied scenarios online. This model is intended to address the uncertainty between CPU utilization and workload as described in Challenge 2. 
 Specifically, our model takes the target CPU utilization $x_{cpu}$ and id feature $x_{id}$ of the app as input $X$, and output the per-Pod workload $y$. Our regression model ($y=w \cdot x_{cpu}+b$), adopts Bayesian neural networks \cite{blundell2015weight} to provide the estimate of the uncertainty. 
The parameters $\beta$ include weight $w$ and bias $b$. We assume parameters $\beta$ follows probabilistic distribution, where the prior distribution is $p_{prior}(\beta) \sim \mathcal{N}(0,I)$, and the posterior distribution is $p_{posterior}(\beta|X,y)$. With the help of the variational Bayes, we can use the learnable model $q_{\phi}(\beta|X,y) \sim \mathcal{N}(\bm{\mu},\bm{\Sigma})$ as an approximation to the intractable true posterior distribution, where $\phi$ is the parameter of the model given by task conditioned information.
 
We first compute a task embedding $I_{\tau}$ for each scenario of the app, using a task projector network $h_{I}(\cdot)$, which is an MLP consisting of two feed-forward layers and a ReLU non-linearity:
 \begin{equation}
      \bm{I_{\tau}}=h_{I}(x_{id})
 \mbox{ ,}
\end{equation}
where $x_{id}$ is the task feature above that indicates different scenarios of the app and the task projector network $h_{I}(\cdot)$ learns a suitable compressed task embedding from input task features.

We then define a simple linear layer of hypernetwork $g_{A}^{l}(\cdot)$, taking the task embeddings $I_{\tau}$ as input, to generate the parameter $\bm{\mu}$(mean) and $\bm{\Sigma}$(variance) of the above approximation distribution of $q_{\phi}(\beta|X,y)$ task-conditioned Bayesian model. In this way, we can provide different model parameters for each estimate of the app.

In the training stage, given $\mathcal{D}$, defined as a batch of data, with the Stochastic Gradient Variational Bayes (SGVB) estimator and reparameterization trick \cite{kingma2013auto}, we train the Task-conditioned Bayesian Neural model following ELBO (Evidence Lower Bound) as:
\begin{small}
\begin{equation}
      \mathcal{L}=
      \frac{1}{|\mathcal D|}(KL[q_{\phi}(\beta|X,y)||p(\beta)]-\mathbb{E}_{\beta \sim q_{\phi}(\beta|X,y)} logP(X,y|\beta))
 \mbox{ .}
\end{equation}
\end{small}
Once the model is trained, we run the model multiple times to generate a set of samples to calculate the mean and variance of the results.  Moreover, by fixing the target CPU utilization, we can determine the corresponding upper and lower bounds of the allocated per-Pod workload.  According to Equation~\eqref{eq:eq1} above, given the total workload $y_{t}^{total}$ and upper and lower bounds of per-Pod workload $y_{t}^{{pod}}$, the range of decision of the number of scaling Pods can be obtained, where the upper bound corresponds to maximized benefits, and the lower bound corresponds to minimized risk. As shown in Figure~\ref{fig:fig1}, we can obtain the overall optimal decision-making scheme to balance benefits and risks (according to service tolerance for \textit{response time}). In practice, our model scales Pods every five minutes to ensure the applications run stably around a target CPU utilization.

\section{Experiment}
In this section, we conduct extensive empirical evaluations on real-world industrial datasets collected from the application servers of our company. Our experiment consists of three parts: \textbf{Workload TS Forecast}, \textbf{CPU-utilization \& Workload Estimation}, and \textbf{Scaling Pods Decision}.

\subsection{Setup}
\noindent\textbf{Datasets.} We collected the 10min-frequency running data of 3000 different online microservices, such as web service, data service, and AI inference service, from our cloud service, including workload, CPU utilization, and the number of Pods.  The workload TS is composed of max RPS (Request Per Second) of applications every 10 minutes. In addition, our data also includes the corresponding CPU utilization and the number of running instances(Pods) every 10 minutes.

\noindent\textbf{Implementations Environment.} All experiments are run on Linux server (Ubuntu 16.04) with Intel$^{\tiny{\textcircled{R}}}$ Xeon$^{\tiny{\textcircled{R}}}$ Silver 4214 2.20GHz CPU, 512GB RAM, and 8 Nvidia$^{\tiny{\textcircled{R}}}$ A100 GPUs.

\subsection{Evalution}

\subsubsection{Workload TS Forecast}
We first pre-train the Multi-scale TS Representation Model with historical data of workload TS. Next, we evaluate our forecasting model with the multi-scale TS Representation.

\paragraph{Baseline and Metrics.}  We conduct the performance comparison against state-of-the-art baselines, including \textit{N-BEATS}, \textit{DeepAR}, \textit{Transformer}, \textit{LogTrans}, \textit{Informer}, \textit{Autoformer}, and \textit{Fedformer}, detailed in Appendix A. We evaluate the workload TS forecast performance via Mean Absolute Error (MAE) and Root Mean Squared Error (RMSE).

\paragraph{Experiment Results.}  We run our experiment 5 times and reported the mean and standard deviation of MAE and RMSE.  The workload TS forecasting results are shown in Table 1, where we can observe that our proposed approach achieves all the best results, with significant improvements in accuracy on real-world high-frequency datasets.

Furthermore, the results of the experiments proved that for high-frequency TS data, the effect of classical deep methods is far inferior to the multi-scale representation fusion model.

\begin{table}
\centering
   \setlength{\tabcolsep}{0.7pt}
  \begin{tabular}{|c|c|c|l|} \hline
  \textbf{Method} & MAE & RMSE \\ \hline
   N-BEATS & 1.851(0.071) & 41.681(0.533) \\ \hline
   DeepAR & 1.734(0.030) & 31.315(0.246) \\ \hline
   Transformer & 1.698(0.018) & 30.359(0.136)\\ \hline
   LogTrans & 1.634(0.013) & 29.531(0.375)  \\ \hline
   Informer & 1.655(0.031) & 30.121(0.679)  \\ \hline
   Autoformer & 1.482(0.043) & 27.489(0.371)\\ \hline
   Fedformer & 1.427(0.075) & 26.761(0.752) \\ \hline
   \textbf{Our model with repr}  
   & $\bm{1.257(0.058)}$ 
   & $\bm{22.763(0.911)}$ 
   \\ \hline
   Our model without repr & 1.873(0.059) & 31.997(0.319)\\ \hline
  \end{tabular}
   \caption{\textbf{The performance (lower is better) averaged over 5 runs.} We conduct the ablation study on our model without repr. }
\end{table}

\paragraph{Ablation Study.}
To further demonstrate the effectiveness of the designs described in Section 3.2, we conduct ablation studies using the variant of our model and the related models. It is obvious that representation-enhanced models outperform vanilla deep models by around 30\%.

As seen in Figure 4,  our TS representation successfully characterizes long-term temporal dependencies across different periods.  We can intuitively see the dependencies between the weekly periods from the representation. Moreover, the TS representation also mines complex nested periods, e.g., the weekly period contains the daily period as shown in the figure.
\begin{figure}[h]
  \centering
  \includegraphics[width=\linewidth]{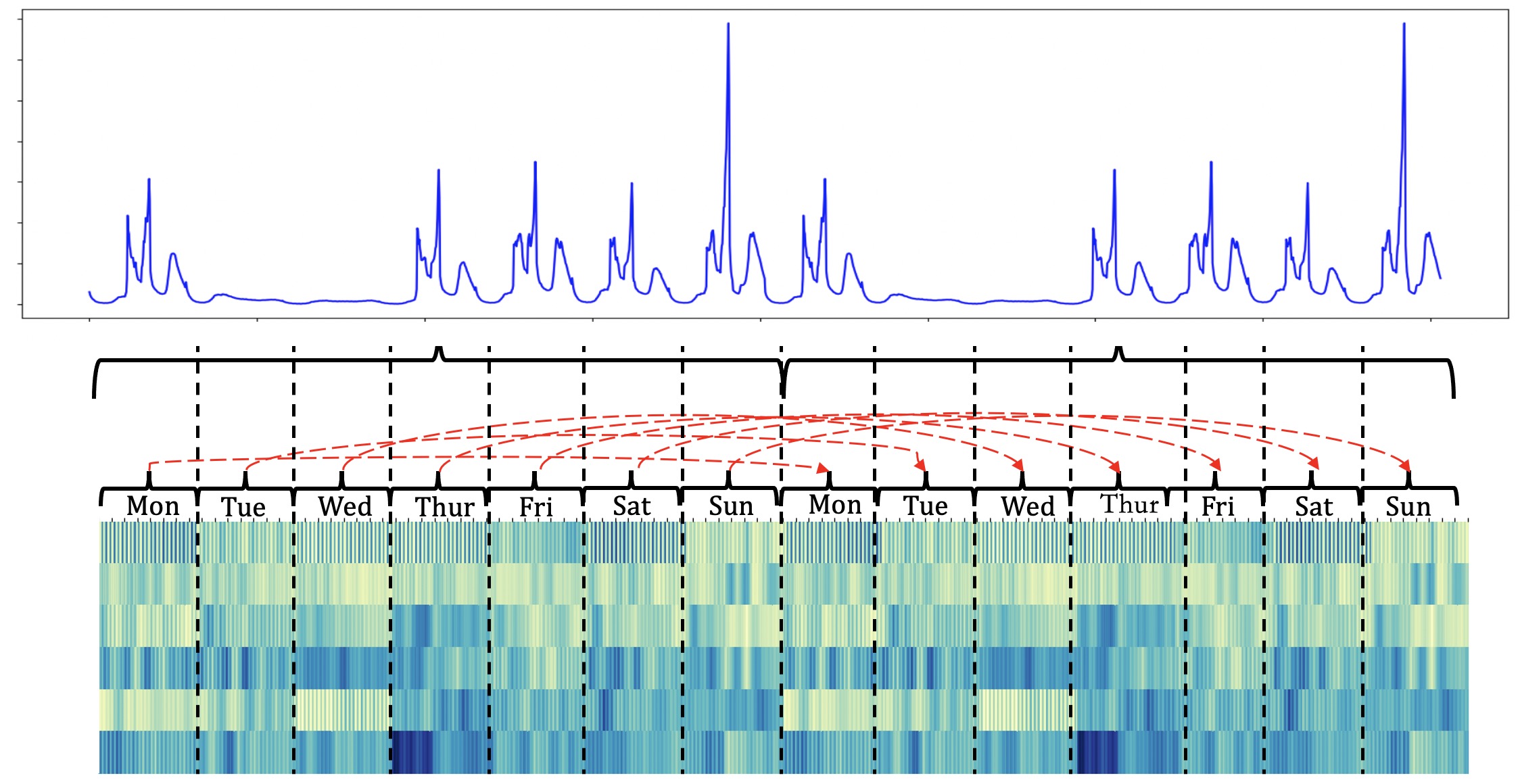}
  \caption{\textbf{Visualization of TS representation of two weeks} (every 10min). Changes in daily and weekly periods are clearly indicated.}
\end{figure}
In summary, benefiting from the powerful TS representation, we can capture workload TS of the high-frequency and non-stationary with various periods and complex long temporal dependencies. Furthermore, with the help of the self-attention-based fusion module, our model integrates long-term historical TS representations and short-term observations (nearby window) for better accuracy.

\subsubsection{CPU-utilization and Workload Estimation}
We take the target CPU utilization and id feature of the app as input, and output the per-Pod workload. 

\paragraph{Baseline and Metrics.}
We conduct the performance comparison against state-of-the-art methods, including \textit{Linear Regression(LR)}, \textit{MLP}, \textit{XGBoost(XGB)}, \textit{Gaussian Process(GP)}, \textit{Neural Process(NP)} \cite{garnelo2018neural}. Please note that for LR, MLP, XGB, and GP, we build independent model for data in each scenario, but in contrast, a globally unified model is used for our approach (theoretically more difficult) and NP (generally considered a meta-learning method) is used to adapt to cross-scenario data, as detailed in Appendix B. We evaluate these experiments via MAE and RMSE.

\paragraph{Experiment Results.}
We run our experiment 5 times and reported the mean and standard deviation of the MAE and RMSE.  Table 2 shows the performances of different approaches on the real-world datasets and we can observe that our proposed approach (Task-conditioned Bayesian Neural Regression) achieves all the best results, with significant improvements in accuracy (around 40\%). Especially, by harnessing the power of the deep Bayesian method, our method characterizes the uncertainty between CPU utilization and workload, providing the upper and lower bounds of benefits and risks. Moreover, we generate different Bayesian model parameters for each set of application data by using a task-conditioned hypernetwork, which allows us to build a globally unified model across scenarios of data, greatly improving modeling efficiency compared with the above traditional methods.

\begin{table}
\centering
  \begin{tabular}{|c|c|c|l|} \hline
  \textbf{Method} & MAE & RMSE \\ \hline
   Linear Regression & 0.712(0.011) & 1.315(0.023) \\ \hline
   MLP & 0.781(0.071) & 1.152(0.533) \\ \hline
   XGBoost & 0.628(0.002) & 1.240(0.004) \\ \hline
   Gaussian Process & 0.677(0.018) &1.271(0.136)\\ \hline
   Neural Process & 0.641(0.013) & 1.275(0.375)  \\ \hline
   \textbf{Our model}  
   & $\bm{0.365(0.005)}$ 
   & $\bm{0.799(0.016)}$ 
   \\ \hline
  \end{tabular}
   \caption{\textbf{The performance (lower is better) averaged over 5 runs. }}
\end{table}

\subsubsection{Scaling Pods Decision}
\begin{figure}[h]
  \centering 
  \includegraphics[width=\linewidth]{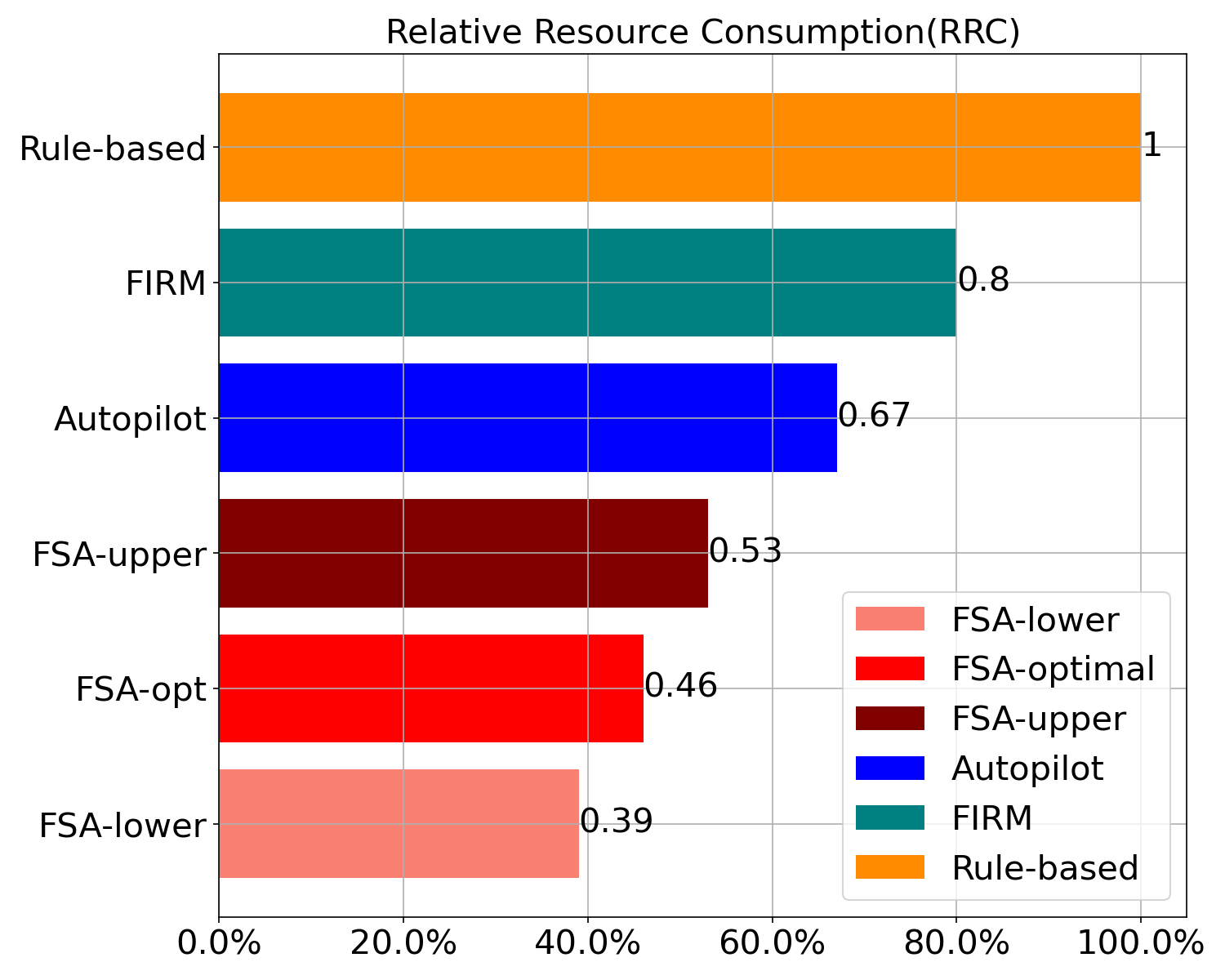}
  \caption{\textbf{The performance(lower is better) of different autoscaling approaches.} The vertical axis represents "Autoscaling method", and the horizontal axis represents "Relative resource consumption-RRC\%".}
  \label{fig:rcc}
\end{figure}

After finishing the above forecast and estimation, 
we obtain the total workload $y_{t}^{total}$ and upper and lower bounds of per-Pod workload $y_{t}^{{pod}}$. Therefore, according to Equation~\eqref{eq:eq1} above, we can obtain the decision range of the number of scaling Pods, where the upper bound corresponds to maximized benefits, and the lower bound corresponds to minimized risk. 

\paragraph{Baseline and Metrics.}
We also conduct the performance comparison against state-of-the-art methods of autoscaling method with the \textit{Rule-based Autoscaling (Rule-based)}, \textit{Autopilot} \cite{rzadca2020autopilot}, \textit{FIRM} \cite{qiu2020firm}. 
We detail the above baselines in Appendix C.

We introduce \textit{relative resource consumption}(\textbf{RRC}) as metrics to measure the performance of autoscaling, defined as:
\begin{equation}
    RRC=c/c_{r}
\mbox{ ,}\end{equation}
where $c$ is the Pod count allocated by the method for evaluation, and $c_{r}$ is the Pod count set by the baseline rule-based autoscaling method. Lower RRC indicates better efficiency (i.e., lower resource consumption).

\paragraph{Experiment Results.}
We run our experiment on 3000 online microservice applications to adjust Pods and report the RRC and scaling Pods count in Figure~\ref{fig:rcc}.

As seen in Figure~\ref{fig:rcc}, FSA demonstrates a significant performance improvement compared to other methods. On the premise of providing stable SLOs assurances and maintaining a target CPU utilization, FSA utilizes the least amount of resources. In particular, FSA provides the range of decisions for scaling Pods, including upper and lower bounds that represent benefit and risk respectively. We also can obtain the optimal value between the upper and lower bound according to service tolerance for \textit{response time} (RT). Figure 5(b) shows the scaling Pods count by running each method in one day, which demonstrates the effectiveness and robustness of FSA. We characterize uncertainty in the decision-making process to provide upper and lower bounds for scaling, which allows our method to achieve stable SLOs assurances.

\begin{figure}[h]
  \centering
  \includegraphics[width=\linewidth]{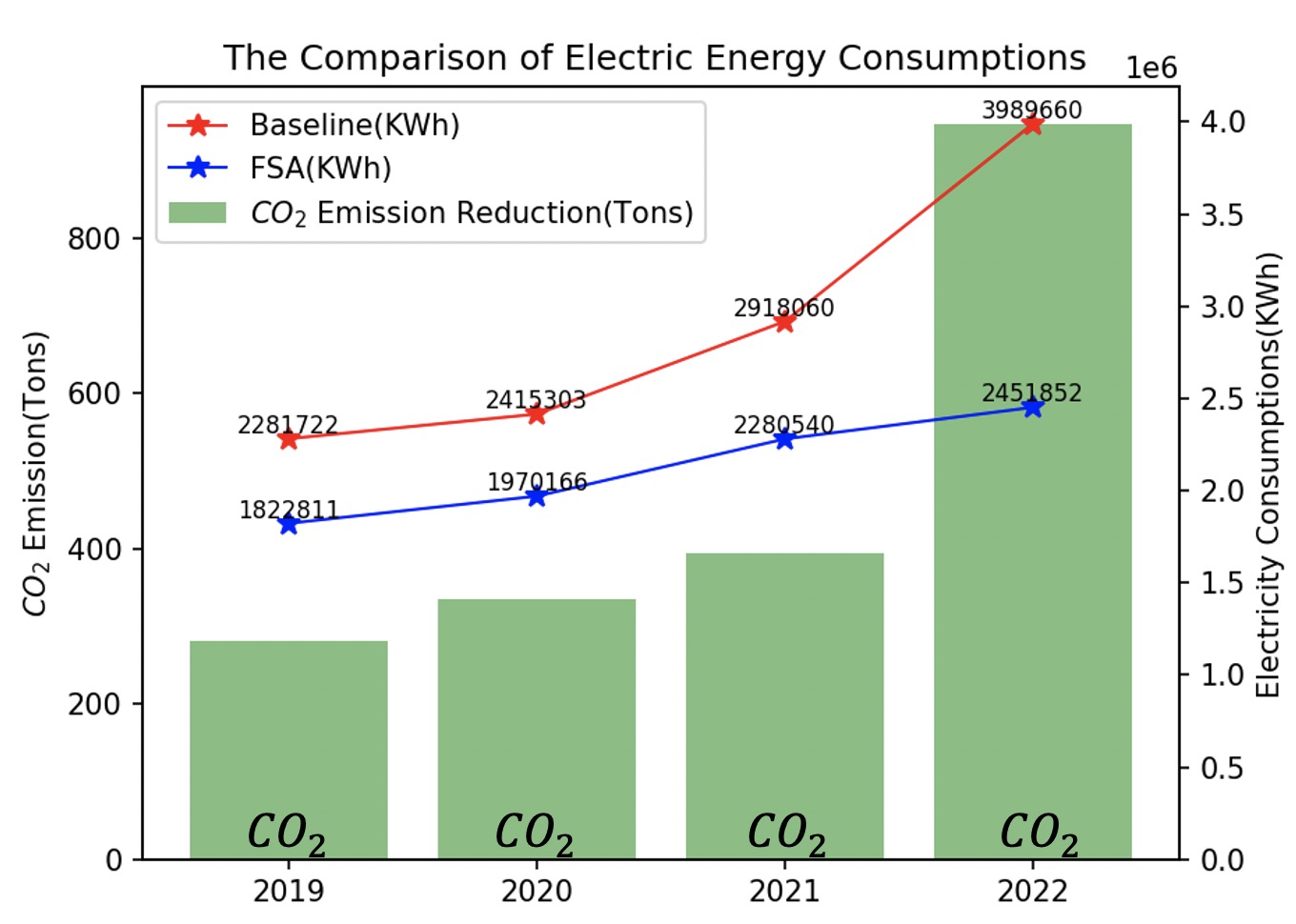}
  \caption{\textbf{The Comparison of Electric Energy Consumptions.} The \textcolor{red}{red} line represents the electricity consumption of baseline, and \textcolor{blue}{blue} line is the electricity cost of FSA. The \textcolor{Green}{green} histogram is the carbon emission reduction using FSA over the past four years. }
\end{figure}

\section{Deployment}

Our approach is successfully deployed on the server cluster in our company's data center, supporting the resource scheduling of more than 3000 online application services. As seen in Figure 6, we report the effectiveness of our approach in enhancing the sustainability of data centers over the past four years, resulting in greater resource efficiency and savings in electric energy.  More details can be found in Appendix D.

FSA was applied in \textit{Double 11 shopping festivals} over the past \textbf{4 years} and according to the certification of \textit{China Environmental United Certification Center} (CEC), the reductions of \textbf{282, 335, 394, 947 tons of carbon dioxide}, respectively, (= savings of \textbf{459000, 545000, 640000, 1538000 kWh of electricity}) were achieved during the shopping festivals.

\section{Conclusion}
In this paper, we proposed a Full Scaling Automation (FSA) mechanism to improve the energy efficiency and utilization of large-scale cloud computing clusters in data centers. The FSA mechanism utilizes forecasts of future workload demands to automatically adjust the resources available to meet the desired resource target level. Empirical tests on real-world data demonstrated the competitiveness of this method when compared to the state-of-the-art methods. Furthermore, the successful implementation of FSA in our own data center has reduced energy consumption and carbon emissions, marking an important step forward in our journey towards \textit{carbon neutrality} by 2030.

\section*{Contribution Statement}
The design, implementation, and deployment were carried out  primarily by Shiyu Wang, Yinbo Sun, and Xiaoming Shi - three authors who played a key role in this work.

\bibliographystyle{named}
\bibliography{ijcai23}

\end{document}